\documentclass{ws-procs9x6}

\begin{document}

\title{Cubic order for spatial 't Hooft loop in hot QCD}

\author{P. Giovannangeli{\footnote{\uppercase{W}ork partially
supported by \uppercase{M}\uppercase{E}\uppercase{N}\uppercase{E}\uppercase{S}\uppercase{R}}}
{\null} and C.P. Korthals Altes}

\address{Centre Physique Theorique au CNRS \\
Case 907, Campus de Luminy, \\ 
F13288, Marseille, France\\ 
E-mail:giovanna@cpt.univ-mrs.fr, altes@cpt.univ-mrs.fr}


\maketitle

\abstracts{Spatial 't Hooft loops of strength $k$ measure the qualitative change in the behaviour of electric colour flux in confined and deconfined phase of SU(N) gauge theory. They show an area law in the deconfined phase, known analytically to two loop order with a ``k-scaling'' law $k(N-k)$. In this paper we compute the $O(g^3)$ correction to the tension. It is due to neutral gluon fields that get their mass through interaction with the wall. The simple k-scaling  is lost in cubic order. The generic problem of non-convexity  shows up in this order. The result for large N is explicitely given.
}

\section{Introduction}
The spatial 't Hooft loop~\cite{thooft} monitors the behaviour of chromoelectric flux. In the confined phase of QCD the flux behaves quite different from that in the deconfined phase. Hence it is an order parameter for gluodynamics, and to some extent also for full QCD~\cite{korthalskovner}. 

It is a perturbatively calculable quantity~\cite{bhatta}. More precisely it  can be computed as a tunneling effect through a potential mountain that is perturbatively calculable. The potential has a periodic structure due to Z(N) symmetry~\cite{thooft}. This Z(N) symmetry shows up in every gauge system with a compact periodic dimension and leads to an area law for the loop at high temperature. Hence the calculation described below is equally valid in for domainwalls one will encounter in a four dimensional world with a compact fifth dimension~\cite{korthalslaine}.

In this note we compute the $O(g^3)$ correction to the tension in the large $N$ limit. 

\section{What is the 't Hooft loop?}
The 't Hooft loop $V_k(L)$ is usually defined~\cite{thooft} as a
closed magnetic flux loop $L$ of strength $k{2\pi\over N}$, with N the number of colours. In operator language it is defined as a gauge transformation with
a discontinuity   $\exp{ik{2\pi\over N}}$ when crossing the minimal surface.
As we will justify below, one can write this gauge transform for $L$ in the x-y plane as a dipole sheet~\cite{korthalskovner} in the following way:

\begin{equation}
V_{k}(L)=\exp{i{4\pi\over N}\int_{S(L)} dxdyTrE_zY_k}.
\label{eq:loop}
\end{equation}
 The $NxN$ diagonal traceless matrix $Y_k$ is defined as
\begin{equation}
Y_k=diag (k,k...k,k-N,k-N.....,k-N)
\label{eq:charge}
\end{equation}
with N-k entries k and k entries k-N, to have a traceless matrix.
The charges $Y_k$ are generalizations of the familiar hypercharge, with k=1.
The charge $Y_k$ of a gluon is 0 or $\pm N$. The multiplicity of the value $N$ is $k(N-k)$.  The same is true for the value $-N$.  So, e.g.  for $N=3$ and $k=1$ one finds the four kaons with hypercharge $\pm 3$.

Exponentiation of  $Y_k$ gives $\exp{i{2\pi\over N}Y_k}=\exp{ik{2\pi\over N}}\equiv z_k$, the centergroup element.
 
$E_z$ is the z component of the canonical electric field strength operator
$\vec E=\lambda_a \vec E$ \footnote{ The $\lambda$ matrices being normalized to 
$Tr\lambda_a\lambda_b={1\over 2}\delta_{a,b}$.}. 


\section{How to compute its thermal average.}

Apart from its obvious connection with colour electric flux the 't Hooft loop
is intimately related to the Z(N) symmetry. Take its order
parameter $P(A_0)$, and move it through the dipole layer. Then it will get
multiplied by the discontinuity $\exp{ik{2\pi\over N}}$, being a  fundamental test charge.
 Immersing the loop in the plasma induces a disturbance. The disturbance is described by a profile $C$. Since the loop is gauge invariant the response of the plasma is too. This profile is 
the phase of the Polyakov loop as a function of its distance to the minimal surface.

So the approach will be to compute the free energy excess $\Delta F$ due to the presence
of the Polyakov loop profile. Let the box be of size $L_{tr}^2*L_z$, with $L_z>>L_{tr}$
and both macroscopic.
Extend the loop to the full x-y cross-section of size $L_{tr}^2$ and located at say $z=0$.

Then we have $\exp{-\Delta F(C)}=\int DC\exp{-{L_{tr}^2\over{g^2}} U(C)}$. $U(C)$ is the constrained effective potential:

\begin{equation}
\exp{-{{L_{tr}^2\over{g^2}}U\big(\widetilde P\big)}\equiv \int DA_0 D\vec A\delta\big(\widetilde P-\bar P(A_0)\big)\exp{{1\over{g^2}}S(A,s_k)}}.
\label{eq:constraint}
\end{equation}
and we have used an abbreviated notation for the constraint:
\begin{equation}
\delta\big(\widetilde P-\bar P(A_0)\big)=\Pi_{z,l}\delta\big(\widetilde P^{(l)}(z)-\bar P^l(A_0(z))\big)
\end{equation}
where $l$ runs from 1 to $N-1$.
In the following we parametrize the fixed loop by a diagonal traceless $NxN$ matrix $C(z)=diag(C_1,C_2,.....C_N)$ :
\begin{equation}
\widetilde P^{(l)}\equiv {1\over N}Tr\exp{ilC(z)}.
\end{equation}
So the effective potential is defined on the Cartan space in which the matrices $C$ live. The matrix describes the profile of the loop.

 It also fixes the 
path in Cartan space (the space of diagonal $C$'s) along which the minimal
profile is realized. 
The path turns out to be the  simplest possible~\cite{giovanna}:
if parametrized by $q$, $0<q<1$, it is given by the one dimensional set
of Cartan matrices 
\begin{equation}
Y_k(q)=qY_k 
\label{eq:path}
\end{equation}
\noindent with $Y_k$ the charge characterizing the 
strength of the dipole layer, eq.(\ref{eq:loop}).
 In exponentiated form it goes from 1 to
$\exp{ik{2\pi\over N}}$, as $q$ goes from $0$ to $1$.
Z(N) invariance of the profile functional $U(C)$ garantuees we we can
take a smooth path from 1 to $\exp{ik{2\pi\over N}}$, instead of a path
that makes a jump $\exp{ik{2\pi\over N}}$ at the loop and returns to 1.
A  proof of the rectilinear path being the minimal one is still lacking.
Only for SU(3) and SU(4) it is known to be the case by inspection~\cite{bhatta}~\cite{giovanna} and a proof at large N is given in ref.~\cite{bhatta}.


\section{Perturbative results}
A loop expansion of the constrained potential eq.(\ref{eq:constraint})
is straightforward. One introduces the colour diagonal background field $B$ through $A_{\mu}=B\delta_{\mu,0}+gQ_{\mu}$.

The saddle point expension around the Polyakov loop $P(\widetilde C)$ fixes the background $B$ in terms of $C$. 

The results from one and two loops are known~\cite{bhatta}~\cite{giovanna} and are given by 
\begin{equation}
\rho_k(L)=k(N-k){4\pi^2\over{3(3g^2N)^{1\over 2}}}T^2(1- 1.1112..(\alpha_s N)+O(g^3))
\end{equation}
The gauge coupling is running according to the $\overline{MS}$ scheme~\cite{recentkaj}.

\section{Three loop potential and self energy matrix}\label{subsec:3looppot}

\begin{figure}
\begin{center}
\includegraphics{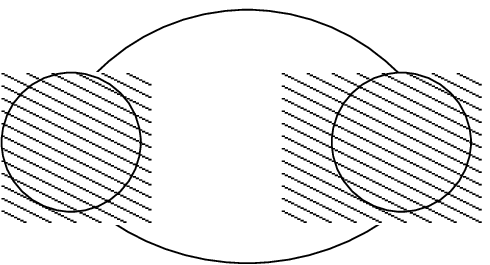}
\caption{The only  three loop diagram of  free energy topology with an infrared divergence. The shaded blob is the one loop selfenergy of. The colour index $c,c'$ of the two propagators need not be the same due to background dependence inside the blobs.}
\label{fig:Df}
\end{center}
\end{figure}

This diagram is in the case of vanishing background linearly divergent for the
Coulombic components of the propagator. 
For notational convenience we write $\Pi_{c,c'}^{(k)}(q)$ for the Coulombic selfenergy matrix at zero momentum and frequency, dropping the Lorentz indices. The index k means we are evaluating the selfenergy along the tunneling path $Y_k(q)$. It is not diagonal except for $k=1$.
The eigenvalues $\Pi^{(k)}_e(q)$~\cite{cubic} turn out to be of the form:
$$1-6{k\over N}(q-q^2)\equiv \big(f(r,q)+1\big)^{2\over 3}$$
So the inverse Coulomb propagator in the background $q$ becomes:
\begin{equation}
\vec l^2\delta_{c,c'}+\Pi_{c,c'}^{(k)}(q)
\label{eq:coulomb}
\end{equation}
and gives an order $g^3$ contribution to the effective potential. In particular for $q=0$, this contribution is proportional to $m_D^3$. 

For the infrared finite result to be Z(N) invariant we should keep 
  contributions from all Matsubara frequencies in 
 the selfenergy.

 Subtract from from all selfenergies in 3 or more loops the zero frequency-momentum piece, to get convergent integarls. Divergencies due to  the magnetic sector remain.

\section{The $O(g^3)$ contribution to the tension.}\label{sec:g3}

We have now eliminated the infrared divergencies from the potential up and including three loop order.
By resumming the Z(N) invariant eigenvalues  this procedure
respects the Z(N) invariance of the potential.

The  price to pay
is a negative  selfenergy   in a certain window of q values of order 1. 
 The problem has not to do with the infrared
scales. It is a generic problem occurring in  quantum corrections to 
tunneling through a barrier.  However,  for a window of loop strengths ${1\over 3}<r={k\over N}<{2\over 3}$ the mass stays positive over the whole range of q values.  In particular in the large N limit we will be able to extract the potential without this problem.

The cubic term in the tension is obtained by minimizing the effective potential eq.(\ref{eq:constraint}).

The tension becomes then to cubic order:
\begin{equation}
\rho_r(T)=\rho_r^{(1)}\big(1- 1.1112..(\alpha_s N)+({3\over{\pi^3}})^{{1\over 2}}I(r)(\alpha_sN)^{{3\over 2}}+O(\alpha_s^2)\big)
\label{eq:finalresult}
\end{equation}
and  
\begin{equation}
I(r)\equiv \int dq{(f(r,q)(1-r)^2+f(1-r,q)r^2)\over {6q(q-1)r(1-r)}}
\label{eq:Ifunction}
\end{equation}
\noindent  As in the pressure~\cite{recentkaj} it contributes with a sign opposite
to the $O(g^2)$ term.
 
The attraction between loops becomes stronger due to the convexity of $I(r)$.
To see this, take the ratio of the tension of the k-loop and compare it to
the k times the tension of the elementary loop with k=1:
\begin{equation}
{\rho_{k\over N}(T)\over{k\rho_{1\over N}(T)}}=(1-r)\big(1+({3\over{\pi^3}})^{{1\over 2}}(I(r)-I(0))(\alpha_sN)^{{3\over 2}}+O(\alpha^2\big)+O({1\over{N^2}})
\end{equation}
The cubic correction has now a negative coefficient, so that the ratio is smaller due to the presence of this correction.

\end{document}